\documentclass[11pt]{article}
\def\K{{K\"{a}hler}}
\newsavebox{\uuunit}
\sbox{\uuunit}
    {\setlength{\unitlength}{0.825em}
     \begin{picture}(0.6,0.7)
        \thinlines
        \put(0,0){\line(1,0){0.5}}
        \put(0.15,0){\line(0,1){0.7}}
        \put(0.35,0){\line(0,1){0.8}}
       \multiput(0.3,0.8)(-0.04,-0.02){12}{\rule{0.5pt}{0.5pt}}
     \end {picture}}

\usepackage{color}

\usepackage[pdftex]{hyperref}
\usepackage{url,cite}

\usepackage{amssymb,slashed,latexsym,ifpdf,amsmath,dsfont}

\newcommand{\rf}[1]{(\ref{#1})}

\newcommand{\be}{\begin{equation}}
\newcommand{\ee}{\end{equation}}
\newcommand{\ba}{\begin{eqnarray}}
\newcommand{\ea}{\end{eqnarray}}

\newcommand{\lp}{\left(}
\newcommand{\rp}{\right)}

\newcommand{\e}{\textrm{e}}

\def\ib{{\bar \imath}}
\def\jb{{\bar \jmath}}
\newcommand{\BS}{{\bf S}}

\newcommand{\BPhi}{{\bf \Phi}}
\newcommand{\BPhibar}{{\bf \bar \Phi}}

\usepackage{color}

\begin{document}

 %%%%%%%%%%%%%%%%%%%%%%%%%%%%%%%%%%%%%%%%%%%%%%%%%%%%%%%%%%%
\begin{titlepage}
\begin{flushright}
TUW-16-03
\end{flushright}
\vspace{.5cm}
\begin{center}
\baselineskip=16pt

%%%%%%%%%%%%%
{\huge {\bf Coupling the Inflationary Sector  to Matter}}

\

\

  {\Large Renata Kallosh$^1$,   Andrei Linde$^1$  and Timm Wrase$^2$} \vskip 0.8cm
{\normalsize\sl\noindent
$^1$ SITP and Department of Physics, Stanford University, Stanford, CA
94305, USA \\\smallskip
$^2$Institute for Theoretical Physics, TU Wien, A-1040 Vienna, Austria}
\end{center}

\vskip 3cm
\begin{abstract}

{\normalsize We describe the coupling of matter fields to an inflationary sector of supergravity,  the inflaton $\Phi$ and a stabilizer $S$, in models where the \K\, potential has a flat inflaton direction.  Such models include, in particular, advanced versions of the hyperbolic $\alpha$-attractor models with a flat inflaton direction \K\, potential,  providing a good fit to the observational data. If the superpotential is at least quadratic in the matter fields $U^{i}$, with restricted  couplings to the inflaton sector, we prove that under certain conditions:  i) The presence of the matter fields does not affect a successful inflationary evolution. ii)~There are no tachyons in the matter sector during and after inflation. iii)~The matter masses squared are higher than $3H^2$ during inflation. The simplest class of theories satisfying all required conditions is provided by models with a flat direction \K\ potential, and with the inflaton $\Phi$ and a stabilizer $S$ belonging to a hidden sector, so that matter fields have no direct coupling to the inflationary sector in the \K\, potential and  in the superpotential.}

 \end{abstract}

 \vspace{2mm} \vfill \hrule width 3.cm
{\small \noindent alinde@stanford.edu, kallosh@stanford.edu, timm.wrase@tuwien.ac.at }
 
 \end{titlepage}
 
%%%%%%%%%%%%%%%%%%%%%%

 \tableofcontents{}
%\newpage

\section{Introduction}\label{intro}

During the last years there was a significant progress in constructing realistic inflationary models in supergravity. This difficult task can be most efficiently achieved in theories with \K\ potentials with a flat direction corresponding to the inflaton field. In such theories, the inflaton potential has a shift symmetry broken only by the superpotential. This allows to develop a large class of flat inflationary potentials suitable for the implementation of various versions of chaotic inflation \cite{Kawasaki:2000yn, Kallosh:2010xz}. 

It was found also that many successful versions of supergravity inflation, which had flat potentials for seemingly unrelated reasons, may in fact be equivalently represented as theories with \K\ potentials with a flat direction. This includes, in particular, the very first version of chaotic inflation in supergravity  \cite{Goncharov:1983mw}. 
More recent progress is related to the development of a class of $\alpha$-attractors \cite{Kallosh:2013hoa,Carrasco:2015uma,Carrasco:2015rva,Carrasco:2015pla} based on the hyperbolic geometry of the moduli space \cite{Kallosh:2015zsa}. These models provide an excellent fit to the latest cosmological data \cite{Ade:2015lrj}. 
Flatness of the inflaton potential in the original formulation of these models did not require flatness of the \K\ potential in the inflaton direction;  is was related to the hyperbolic geometry of the moduli space \cite{Kallosh:2015zsa}. However, using the symmetries of the hyperbolic geometry it was possible to cast the original \K\ potentials of these models into a form with the flat direction corresponding to the inflaton field \cite{Carrasco:2015uma,Carrasco:2015rva}. This helped to solve the problem of initial conditions for inflationary models of this type \cite{Carrasco:2015rva,East:2015ggf}, and facilitated the development of models which could simultaneously describe inflation, dark energy and supersymmetry breaking \cite{Carrasco:2015pla}. The formulation of such models is especially simple in the context of the theory of orthogonal nilpotent superfields \cite{Ferrara:2015tyn,Carrasco:2015iij,Dall'Agata:2015lek}, where the \K\ potential is  flat in the inflaton direction.

In this paper we will reveal yet another advantage of inflationary models with  \K\, potentials with a flat direction: A broad class of such models, including all models mentioned above, can be consistently generalized by adding matter multiplets. Under certain conditions to be specified below, the  scalar fields from the matter multiplets have all vanishing vacuum expectation values during inflation, and therefore they do not affect the inflationary dynamics.  This is rather important, because in general one could expect that interactions between various fields could lead to tachyonic instabilities and a very complicated multi-field cosmological evolution. Meanwhile in the theories to be discussed below one can introduce matter multiplets while preserving all advantages of the previously constructed inflationary models.

More specifically, we will consider inflationary models with the inflaton $\Phi$ and a stabilizer $S$. The stabilizer $S$ can be either a  nilpotent superfield $S^2(x, \theta)\equiv\BS^2=0$ or just a very heavy field due to the presence of the sectional curvature term  $- (S\bar S)^2/\Lambda^2$ in the \K\, potential \cite{Kallosh:2010xz}. We assume that the K\"ahler potential for the matter fields $U^i$ is canonical, and that there is no direct coupling in the \K\ potential between the two sectors: 
\be
K= K_{\rm infl}(\Phi, \bar \Phi; S, \bar S)  + K_{\rm mat} (U^i, \bar U^\ib)= K_{\rm infl}(\Phi, \bar \Phi; S, \bar S)  +\sum_i U^i \bar U^\ib \ .
\ee
The simplest option to consider is that the inflaton field and the stabilizer belong to the hidden sector, so there is no direct interactions between these fields and the matter fields $U^{i}$, not only in the \K\ potential but also  in the superpotential. This separation can be very useful, if one wants to suppress the reheating temperature in the theory. However, our main results will be valid in a more general class of theories, so we will start with a more general superpotential which can describe direct interactions between all fields, 
\be
W= W_{\rm infl} (\Phi, S)+ W_{\rm mat}(U^i, \Phi, S) \ ,
\label{Wdecoupled}\ee
where
\be
W_{mat}(U^i, \Phi, S)={1\over 2}  A_{ij} U^i U^j + {1\over 3!}B_{ijk} U^i U^j U^k + \ldots
\label{quadratic}\ee
Here $\ldots$ denotes terms with higher powers in the $U^i$ and $ A_{ij}$, $B_{ijk}$, etc. may, in principle, depend on the inflaton sector. But to guarantee the absence of tachyons, we will find that such a dependence in $A_{ij}$ is very restricted.

Our condition that the matter multiplets have vanishing vev's and that the \K\, potential and superpotential are quadratic in matter fields is a choice valid, for example, for the superfields like squarks and leptons
\be
U^i=0\,.
\ee
In this case we are able to establish rather general conditions for the absence of tachyons. The case of matter fields with non-vanishing vev's has to be investigated separately. See also Appendix \ref{higgs} for a discussion of these issues.

Various studies of multi-sector supergravities were performed in the past. For example cases with $K= K^1 + K^2$ and $W= W^1\cdot W^2$ were studied  in \cite{Achucarro:2011yc} where also the references to earlier work are given. This is different from our setting in 
eq. \rf{Wdecoupled}.
In inflationary models developed in \cite{Dall'Agata:2014oka} it was proposed to have a superpotential at least quadratic in matter fields, however, due to the nature of these models, the coupling of matter fields to the nilpotent superfield was required to be of the form $W(\Phi, S, U^i) = f^{\rm DZ}(\Phi, U^i) (1+\sqrt 3 S)$. This is a subclass of our models considered in eq.~\rf{Wdecoupled}. 

We will find that in general our models in eqs. \rf{Wdecoupled} and \rf{quadratic} might have some tachyonic  matter directions during inflation. However, the situation becomes much better, if the inflaton field corresponds to a flat direction of the \K\ potential. For definiteness, we will assume that the inflaton direction corresponds to the real  field   $\Phi+\bar \Phi$ and 
the \K\, potential in the inflaton direction $\Phi+\bar \Phi$ is flat\footnote{ We will often refer to these models which have a flat direction in a \K\, potential as models `with a flat \K\, potential'. Note, however, that this should not be confused with the \K\, manifold curvature. For example in $\alpha$-attractor models \cite{Carrasco:2015uma,Carrasco:2015rva,Carrasco:2015pla} the curvature is ${\cal R}_K= -{2\over 3 \alpha}$, but the \K\, potential has a flat inflaton direction.}, i.e.
\be
\partial_\Phi K\Big |_{\Phi-\bar \Phi=0}=0\, ,
\label{flat}\ee
whereas the sinflaton $\Phi-\bar \Phi$ is stabilized at $\Phi-\bar \Phi=0$.
In such a case we find that all tachyons disappear under the condition that $A_{ij}$ is either independent of $\Phi, S$ or has a certain restricted dependence on the inflaton sector fields. The simplest \K\ potential of such type is $-(\Phi-\bar\Phi)^{2}/2$  \cite{Kallosh:2010xz}. Magically,  the hyperbolic $\alpha$-attractor models \cite{Carrasco:2015uma,Carrasco:2015rva,Carrasco:2015pla}, compatible with the current and future data, belong to this class of `{\K} flat' models.

A more general class of  \K\, potentials described in \cite{Kallosh:2010xz} also satisfies this condition. It requires the \K\, potential to be invariant under the following transformations
\be
\Phi \leftrightarrow \bar \Phi\, , \qquad 
\Phi \rightarrow \Phi +a,  \qquad a \in \mathbb{R}\,.
\ee
In many of such models, the inflationary trajectory $\Phi-\Bar \Phi=0$ is a stable minimum with respect to the field $\Phi-\Bar \Phi$. 
The requirement that inflation takes place at $\Phi-\Bar \Phi=0$ can be also implemented by either using the nilpotent orthogonal superfields $\BS(\BPhi-\BPhibar)=0$ as proposed in  \cite{Ferrara:2015tyn,Carrasco:2015iij}, or, if necessary, by adding a sinflaton stabilization term to the \K\, potential of the form $A\,  S \bar  S(\Phi-\bar \Phi)^2$  \cite{Kallosh:2010xz}. Here $A$ defines the bisectional curvature of the \K\, manifold which may depend on the inflaton field. 

The condition for inflation to take place at $\Phi-\Bar \Phi=0$ usually requires the superpotential to be a holomorphic function of the superfields with real coefficients, as explained in \cite{Kallosh:2010xz,Kallosh:2014xwa}. This condition may be relaxed in theories with nilpotent orthogonal superfields  \cite{Ferrara:2015tyn,Carrasco:2015iij,Dall'Agata:2015lek} where $\Phi-\Bar \Phi=0$ by construction.
Also, as we will show in this paper,  in the theories with  nilpotent orthogonal superfields one can introduce matter fields with arbitrary dependence of $A_{ij}$ on the inflaton field, without affecting stability of the inflationary trajectory.

Note that one may equivalently  consider \K\ potentials which are flat in the imaginary direction, such as $(\Phi+\bar\Phi)^{2}/2$ \cite{Kawasaki:2000yn,Dall'Agata:2014oka}. In that case the formulation of the condition equivalent to the reality condition mentioned above is slightly more involved, see \cite{Kallosh:2014xwa}.

%%%%%%%%%%%%%%%%
\section{Coupling the inflationary sector to matter}

\subsection{General case}\label{sec:general}

We start with a model of inflation defined via two fields $\Phi$ and $S$ and the following K\"ahler and superpotential
\be\label{eq:inf}
K=k(\Phi,\bar \Phi) +S \bar S\,, \qquad W=g(\Phi) + S f(\Phi)\,.
\ee
We will consider nilpotent fields $S^2=0$,\footnote{The same conclusion can be reached for models in which $S$ is not nilpotent, if one adds the sectional curvature term $- (S\bar S)^2/\Lambda^2$ to the \K\, potential and takes $\Lambda$ to be sufficiently small to decouple $S$.} which only allows for an $S$ independent term and a term linear in $S$ in the superpotential.

Now we couple this model to matter fields $U^i$ that satisfies two criteria: 1) The K\"ahler potential is canonical $K_{mat} = \sum_i U^i \bar U^\ib$ and 2) the superpotential $W_{mat}(S,\Phi,U^i)$ is at least quadratic in the fields $U^i$.\footnote{These conditions seem to be satisfied for the MSSM and NMSSM. We also only need the weaker condition that $U^i=0$ implies $\partial_i K=\partial_{ij} K=0$.} The full model is then
\ba\label{fullmodel}
K&=&k(\Phi,\bar \Phi) +S \bar S+\sum_i U^i \bar U^\ib\,,\cr
W&=& g(\Phi) + S f(\Phi) + W_{mat}^g(\Phi,U^i) + S W_{mat}^f(\Phi,U^i)\,.
\ea
Here $W_{mat}^g$ adds a matter dependent term to $g(\Phi)$ and $W_{mat}^f$ adds a matter dependent term to $f(\Phi)$. In general, we will study the case where the term in the superpotential that is quadratic in the $U^i$ depends on $\Phi$ and $S$:
\be
W= g(\Phi) + S f(\Phi) + {1\over 2} \Big (A^g_{ij}(\Phi) + S A^f_{ij}(\Phi)\Big) U^i U^j + \ldots
\label{coupling}\ee
i.e. we specify that in \rf{quadratic} $A_{ij}(\Phi,S) = A^g_{ij}(\Phi) + S A^f_{ij}(\Phi)$.

We find the following scalar potential
\ba
V&=&e^{k+\sum_i U^i \bar U^\ib} \lp|f+W_{mat}^f|^2-3|g+W_{mat}^g|^2 +\sum_i |\partial_i W_{mat}^g + \bar U^\ib(g+W_{mat}^g)|^2 \right.\cr
&&\quad \qquad \qquad \left.+K^{\Phi\bar \Phi} |g' +(W_{mat}^g)'+(g+W_{mat}^g) \partial_\Phi k|^2 \rp   .
\ea
Here $'$ means $\partial_\Phi$ on holomorphic functions and $\partial_{\bar \Phi}$ on anti-holomorphic functions.
Thanks to the above requirements, we see that for 
\be\label{Uzero}
U^i=0 \quad \text{we have} \quad W_{mat}^{g/f}=\partial_i W_{mat}^{g/f}=\partial_\Phi^n W_{mat}^{g/f}=\partial_\Phi^n \partial_i W_{mat}^{g/f}=0\,, \quad \forall n \,,
\ee
and the scalar potential in this case reduces to
\be
V_{inf}=e^{k} \lp|f|^2-3|g|^2 +K^{\Phi\bar \Phi} |g'+g \partial_\Phi k|^2 \rp\,,
\ee
which is just the inflaton potential resulting from \eqref{eq:inf}.

The interesting fact is that $U^i=0$ is actually a critical point at which the mass matrix is block diagonal with one block corresponding to the inflaton sector and the other block corresponding to the matter sector. 

Let us first show that $U^i=0$ is a critical point:
\ba
\partial_i V &=& \bar{U}^\ib V + e^{k+\sum_l U^l \bar U^{\bar l}} \Big[(\partial_i W_{mat}^f)(\overline{f+W_{mat}^f}) -3 \partial_i W_{mat}^g (\overline{g+W_{mat}^g}) \cr
&&+ \sum_j(\partial_{ij} W_{mat}^g + \bar U^\jb \partial_i W_{mat}^g)(\overline{\partial_j W_{mat}^g}+  U^j (\overline{g+W_{mat}^g})) \cr
&& +(\partial_i W_{mat}^g + \bar U^\ib(g+W_{mat}^g))(\overline{g+W_{mat}^g})\cr
&&+K^{\Phi\bar \Phi} (\partial_i (W_{mat}^g)'+\partial_i W_{mat}^g \partial_\Phi k) (\overline{g'+(W_{mat}^g)'+(g+W_{mat}^g) \partial_\Phi k}) \Big)\,.
\ea
We see that the above expression vanishes for $U^i=0$ due to eq. \eqref{Uzero}.

We can see analogously that $\partial_\Phi V =0$ reduces for $U^i=0$ to $\partial_\Phi V_{inf} =0$. Similarly, the second derivatives with respect to $\Phi$ and/or $\bar\Phi$ are unchanged and are the same for $V_{inf}$ and $V$ at $U^i= 0$. This follows from eq. \eqref{Uzero}.

Now let us look at mixed derivatives
\ba
\partial_\Phi \partial_i V &=& \bar{U}^\ib \partial_\Phi V + \partial_\Phi\Big[ e^{k+\sum_l U^l \bar U^{\bar l}} \Big((\partial_i W_{mat}^f)(\overline{f+W_{mat}^f})-3 \partial_i W_{mat}^g (\overline{g+W_{mat}^g}) \cr
&&+ \sum_j(\partial_{ij} W_{mat}^g + \bar U^\jb \partial_i W_{mat}^g)(\overline{\partial_j W_{mat}^g}+  U^j (\overline{g+W_{mat}^g})) \cr
&&  +(\partial_i W_{mat}^g + \bar U^\ib(g+W_{mat}^g))(\overline{g+W_{mat}^g})\cr
&&+K^{\Phi\bar \Phi} (\partial_i (W_{mat}^g)'+\partial_i W_{mat}^g \partial_\Phi k) (\overline{g'+(W_{mat}^g)'+(g+W_{mat}^g) \partial_\Phi k}) \Big)\Big]\,.
\ea
We see that every term will be multiplied by a term that vanishes for $U^i=0$ (see eq. \eqref{Uzero}). So we have shown that $U^i=0$ implies $V_{\Phi U^i}=0$. Analogously, we can see that $V_{\bar\Phi U^i}=0$ for $U^i=0$. {\it So we can conclude that the matter sector does not affect the inflationary sector at all.}

Now we check how the inflaton sector affects the matter sector:
\ba
\partial_i \partial_{\jb}V &=& \delta_{i\jb} V + \bar{U}^\ib \partial_{\jb} V + U^j (\partial_i V - \bar U^\ib V) \cr
&&+e^{k+\sum_l U^l \bar U^{\bar l}} \Big((\partial_i W_{mat}^f)(\overline{\partial_j W_{mat}^f})-3 \partial_i W_{mat}^g \overline{\partial_j W_{mat}^g} \cr
&&+ \partial_i W_{mat}^g(\overline{\partial_j W_{mat}^g}+  U^j (\overline{g+W_{mat}^g}))\cr
&&+\sum_l(\partial_{il} W_{mat}^g + \bar U^{\bar l} \partial_i W_{mat}^g)(\overline{\partial_{jl} W_{mat}^g}+  U^l \overline{\partial_j W_{mat}^g}) \cr
&& +(\partial_i W_{mat}^g + \bar U^\ib(g+W_{mat}^g))\overline{\partial_j W_{mat}^g} + \delta_{i\jb}|g+W_{mat}^g|^2\cr
&&+K^{\Phi\bar \Phi} \Big((\partial_i W_{mat}^g)'+\partial_i W_{mat}^g \partial_\Phi k\Big)  \Big(\overline{  (\partial_j W_{mat}^g)'+\partial_j W_{mat}^g \partial_\Phi k} \Big)\,.
\ea
This simplifies substantially for $U^i=0$, where we are left with
\ba
\partial_i \partial_{\jb}V &=& \delta_{i\jb} V +e^{k} \Big( \sum_l(\partial_{il} W_{mat}^g )(\overline{\partial_{jl} W_{mat}^g})+ \delta_{i\jb}|g|^2 \Big) \cr
&=& \delta_{i\jb} \lp V+e^k|g|^2 \rp + e^{k}  \sum_l(\partial_{il} W_{mat}^g )(\overline{\partial_{jl} W_{mat}^g})\,.
\ea
So we find a positive contribution from the inflaton sector to the diagonal entries of the mass matrix.

Let us now check the derivatives along two holomorphic matter directions
\ba
\partial_i \partial_{j}V &=& \bar U^\ib  \partial_j V +  \bar U^\jb (\partial_i V - \bar U^\ib V) +e^{k+\sum_l U^l \bar U^{\bar l}}\Big[(\partial_{ij} W_{mat}^f)(\overline{f+W_{mat}^f})-3 \partial_{ij} W_{mat}^g (\overline{g+W_{mat}^g}) \cr
&&+ \sum_l(\partial_{ijl} W_{mat}^g + \bar U^{\bar l} \partial_{ij} W_{mat}^g)(\overline{\partial_l W_{mat}^g}+  U^l (\overline{g+W_{mat}^g})) \cr
&& +  (\partial_{ij} W_{mat}^g + \bar U^{\jb} \partial_{i} W_{mat}^g)(\overline{g+W_{mat}^g})+(\partial_{ij} W_{mat}^g + \bar U^\ib \partial_j W_{mat}^g)(\overline{g+W_{mat}^g})\cr
&&+K^{\Phi\bar \Phi} \Big((\partial_{ij} W_{mat}^g)'+(\partial_{ij} W_{mat}^g) \partial_\Phi k\Big)  \Big(\overline{g'+(W_{mat}^g)'+(g+W_{mat}^g) \partial_\Phi k} \Big)\,.
\ea
This again simplifies substantially for $U^i=0$ where we are left with
\be
\partial_i \partial_{j}V = e^{k} \lp (\partial_{ij} W_{mat}^f) \bar f + K^{\Phi\bar \Phi} \lp (\partial_{ij} W_{mat}^g)'+(\partial_{ij} W_{mat}^g) \partial_\Phi k \rp(\overline{g'+g \partial_\Phi k}) -(\partial_{ij} W_{mat}^g) \bar g\rp\,.
\ee
The matter mass matrix at $U^i=0$ has thus the following form
\ba\label{matrix}
M^2 &=& \lp\begin{array}{cc}
\partial_{i\jb} V & \partial_{ij} V \\
\partial_{\ib \jb} V & \partial_{\ib j}V
\end{array}\rp\cr
&=& \mathds{1} \lp V+e^k|g|^2 \rp + e^k \lp\begin{array}{cc}
(A^g\cdot \bar A^g)_{i\jb} & c_1 A^g_{ij} + c_2 (A^g_{ij})'+ \bar f A^f_{ij}\\
\bar c_1 \bar A^g_{\ib \jb} +\bar c_2 (\bar A^g_{\ib \jb})' + f \bar A^f_{\ib\jb}& (\bar A^g \cdot A^g)_{\ib j}
\end{array}\rp\,,
\ea
where we defined 
\be
c_1=K^{\Phi\bar \Phi} \partial_\Phi k \, (\overline{g'+g \partial_\Phi k}) -\bar g\,,
\ee
and 
\be
c_2= K^{\Phi\bar \Phi}\, (\overline{g'+g \partial_\Phi k})
\ee
 and we used that $A^{f/g}_{ij} = \partial_{ij} W_{mat}^{f/g}$ for $U^i=0$ (see eq. \eqref{coupling}). \footnote{Note that $B_{ijk}(\Phi,S)$ that controls the couplings that are cubic in the $U^i$ (see eq. \eqref{quadratic}) does not appear at all in the mass matrix and therefore does not affect the stability.}

In general $A^g_{ij}$, $(A^g_{ij})'$ and $A^f_{ij}$ are independent and it is not possible to diagonalize the mass matrix \eqref{matrix}. So we will now discuss more specific cases. In particular, one can diagonalize the mass matrix above, whenever $c_2 (A^g_{ij})'+ \bar f A^f_{ij}=0$ or more generally, when $c_2 (A^g_{ij})'+ \bar f A^f_{ij} \propto  A_{ij}^g$. In this case we define $c$ via
\be
c_1 A^g_{ij} + c_2 (A^g_{ij})'+ \bar f A^f_{ij} \equiv c A^g_{ij}\,,
\ee
and the mass matrix takes the form
\ba\label{matrix2}
M^2 &=& \mathds{1} \lp V+e^k|g|^2 \rp + e^k \lp\begin{array}{cc}
(A^g\cdot \bar A^g)_{i\jb} & c A^g_{ij} \\
\bar{c} \bar A^g_{\ib \jb}& (\bar A^g \cdot A^g)_{\ib j}
\end{array}\rp\,.
\ea
We are thus left with $c A^g_{ij}$ and its conjugate as the off-diagonal terms. To find the eigenvalues of this $M^2$ matrix we do a Takagi factorization \cite{Takagi}, i.e. we write $A^g = U \Sigma U^T$ where $U$ is a unitary matrix whose columns are orthonormal eigenvectors of $A^g\cdot \bar A^g$, $\Sigma$ is a diagonal matrix whose entries $\lambda_1, \ldots, \lambda_n$ are real and non-negative with $\lambda_i^2$ being the eigenvalues of $A^g\cdot \bar A^g$. Now we can perform a unitary transformation and rearrange the rows and columns to bring the matrix into the following block diagonal form
\ba
&&\lp\begin{array}{cc}
U^\dagger & 0 \\
0 & U^T
\end{array}\rp
\lp\begin{array}{cc}
(A^g\cdot \bar A^g)_{i\jb} & c A^g_{ij} \\
\bar c \bar A^g_{\ib \jb} & (\bar A^g \cdot A^g)_{\ib j}
\end{array}\rp
\lp\begin{array}{cc}
U & 0 \\
0 & U^*
\end{array}\rp\cr
\cr
\cr
&
=&
\lp\begin{array}{cc}
\Sigma \cdot \Sigma & c \Sigma \\
\bar c \Sigma & \Sigma \cdot \Sigma
\end{array}\rp
\rightarrow 
\lp\begin{array}{ccccc}
\lambda_1^2 & c \lambda_1&0&0& \\
\bar c \lambda_1 & \lambda_1^2 &0 &0&\\
0&0&\lambda_2^2& c \lambda_2 &\\
0&0&\bar c \lambda_2 & \lambda_2^2&\\
&&&&\ddots
\end{array}\rp
\ea
Now it is trivial to determine the eigenvalues of the mass matrix. We find the following eigenvalues of the matrix  $M^2$ in eq. \rf{matrix2}
\be
 \mu_{i}^{2}= V + \e^{k} |g|^2 +\e^k \lambda_i \lp \lambda_i \pm |c|\rp\,.
\label{general}\ee
Let us discuss the constraint $c_2 (A^g_{ij})'= \bar f A^f_{ij} =0$ in more detail (see subsection \ref{DZ} below for a class of models in which $c_2 (A^g_{ij})'+ \bar f A^f_{ij} \propto  A_{ij}^g$):\\
In order to have $c_2 (A^g_{ij})'=K^{\Phi\bar \Phi} (\overline{g'+g \partial_\Phi k}) (A^g_{ij})'=0$, we need either that the leading matter coupling to the inflaton is absent, in which case the function $g(\Phi)$ can be arbitrary
\be\label{eq:1}
\qquad \partial_\Phi A^g_{ij}(\Phi)= 0 \, , \qquad {\rm arbitrary} \, \, g(\Phi) \qquad \Rightarrow \quad c_2 (A^g_{ij})'=0\,,
\ee
or, if we want the matter coupling $A^g_{ij}(\Phi) U^i U^j$ in the superpotential (see \eqref{coupling}), we need $c_2=0$ 
\be \label{eq:2}
\partial_\Phi A^g_{ij}(\Phi) \neq 0\,, \qquad g'+g \partial_\Phi k=0  \qquad \Rightarrow \quad c_2 (A^g_{ij})'=0 \,.
\ee
This later case is for example realized in models with constant $g$ and a flat {\K} potential (see eq. \eqref{flat}) or in models with vanishing $g(\Phi)$ that we discuss further below in subsection \ref{gzero}.

The other off-diagonal term in the mass matrix, $\bar f A^f_{ij}$, vanishes when either $f=0$ or $A^f_{ij} =0$. If we want to use the nilpotent field $S$ in the inflationary sector, i.e. if we want $f\neq 0$, then we require 
\be
\bar f A^f_{ij}=0 \,, \qquad f(\Phi) \neq 0 \qquad \Rightarrow \quad A^f_{ij}=0\,,
\label{Scoupling}\ee
i.e. there is no coupling of $S$ to the matter fields $U^i$ at the quadratic level in the superpotential in \rf{coupling}. Whenever such couplings are present and $f(\Phi)\neq 0$, we cannot calculate the mass eigenvalues in full generality and we cannot guarantee the absence of tachyons.

\subsection{Stabiliziation using moduli space curvature}\label{curvature}\label{sec:curv}
During inflation the curvature of the moduli space enforces the vanishing of a certain modulus by creating a wall in the potential for a given modulus. At the level of the masses this effect comes from the fact that
the diagonal holomorphic-antiholomirphic part of the mass matrix has a correction due to the curvature of the following form
 \ba\label{eqn:holomorphic-antiholomirphic1}
\Delta{\cal M}_{i\jb}&= & -R_{i\jb k \bar l} \bar F^k  F^{\bar l}  
 \ .
\label{massHAH}\ea
where $F_i\equiv e^{K/2}(\partial_i W+K_i W$), $F^{\ib}= g^{\ib j} F_j$ and  $\bar F^{i}= g^{i \jb } \bar F_{\jb}$.

In general, the curvature of the moduli space is related to the fourth derivative of the \K\, potential as follows
\be
R_{i \jb k \bar l}= {\cal K}_{i \jb k \bar l} - \Gamma^m_{ik} g_{m\bar m} \Gamma^{ \bar m}_{\jb\bar l}.
\ee
The typical choice of the curvature as a stabilizer is when all four directions  in $R_{i \jb k \bar l}$ are such that the corresponding moduli vanish during inflation. The curvature tensor in such a case coincides with the fourth derivative of the \K\, potential since the Christoffel symbols vanish because they are related to a third derivative of the \K\, potential and are therefore linear in one vanishing moduli
\be
R_{i \jb k \bar l}= {\cal K}_{i \jb k \bar l} \, ,  \qquad  \Gamma^m_{ik} =0 \,.
\ee
In the inflationary sector consisting of two superfields $\Phi$ and $S$ curvature stabilizers were proposed in \cite{Kallosh:2010xz}. For example it is known that the negative sectional curvature of the moduli space
$
R_{S\bar S S\bar S} 
$
forces the scalar field field $S$ to reach the minimum at $S=0$ under the condition that $\bar F^S $ and $F^{\bar S}$ are both non-vanishing. 
A bisectional curvature of the kind
$
R_{S\bar S \Phi\bar \Phi}
$
can be constructed to enforce the condition that $\Phi=\bar \Phi$ during inflation.
Such curvature stabilizers  are implemented via the following choice of the non-linear terms in the \K\, potential
\be
\Delta K_S = - A(\Phi, \bar \Phi) (S \bar S)^2 \qquad  \Rightarrow \qquad - R_{S\bar S S\bar S}  \sim A(\Phi, \bar \Phi)
\ee
\be
 \Delta K_{\Phi-\bar \Phi} = - B(\Phi, \bar \Phi) S\bar S (\Phi-\bar \Phi)^2 \qquad   \Rightarrow \qquad - R_{S\bar S \Phi\bar \Phi} \sim B(\Phi, \bar \Phi)
\ee
It seems in principle always possible to make the masses of the matter fields positive, if the underlying {\K} manifold has a corresponding bisectional curvature and if during inflation these matter fields vanish. In particular, if we add to the {\K} potential the term 
\be
\Delta K_U = -\zeta \sum_i S \bar S U^i \bar U^\ib
\ee
then this does not affect any of our conclusions except that the matter masses squared get shifted by $\zeta$: $\mu_i^2 \rightarrow \mu_i^2 + \zeta |f(\Phi)|^2$. So for positive and sufficiently large $\zeta$ all matter fields will have a positive mass squared. We will not consider such contributions here but rather discuss the conditions for the absence of tachyons in the models described by eq. \eqref{fullmodel}.

\subsection{Cases with flat K\"ahler potential and restricted couplings: $c_2 (A^g_{ij})'= \bar f A^f_{ij}=0$}\label{Kflat}

We look for models \rf{fullmodel} where during inflation  $\partial_\Phi K=0$ at $\Phi=\bar \Phi$, which includes our hyperbolic geometry models with shift symmetric K\"ahler potential  \cite{Carrasco:2015uma,Carrasco:2015rva,Carrasco:2015pla}  
\be
k(\Phi, \bar \Phi)= -{3\over 2}   \alpha \log \left[{(1- \Phi\bar\Phi)^2\over (1-\Phi^2) (1-\overline \Phi^2)}  \right] %+S\bar S
\,  ,
\label{KdiskNewDisk}\ee
as well as the simple case 
\be
k(\Phi,\bar \Phi)=  -{1\over 2} (\Phi-\bar \Phi)^2\,.
\ee
A new addition to the class of inflationary models based on orthogonal nilpotent multiplets in \cite{Ferrara:2015tyn,Carrasco:2015iij} also satisfies the flatness condition \rf{flat} of the \K\, potential. Also a general class of models studied in \cite{Kallosh:2010xz} adds more examples in which the condition \rf{flat} is satisfied.

In all these models we also find that during inflation $\e^K=1$ at $\Phi=\bar \Phi$. In such case the mass formula \rf{general} simplifies, since $c=c_1=K^{\Phi\bar \Phi} \partial_\Phi k \, (\overline{g'+g \partial_\Phi k}) -\bar g=-\bar g$, to  \be
\mu_i^2= V + |g|^2 + \lambda_i \lp \lambda_i \pm | g|\rp\,.
\label{masses}\ee
This can be rewritten in the form
\be
\mu_i^2 = V + {3\over 4} |g|^2 + \Big |  \lambda_i \pm {1\over 2} | g|\Big |^2  > 0\,,
\label{Eigenvalues}\ee
or in the form
\be
\mu_i^2 = V + {3\over 4} \lambda_i^2 + \Big |{1\over 2}  \lambda_i \pm  | g|\Big |^2  > 0\,.
\label{Eigenvalues*}\ee
It is important here that the $\lambda_i$ are real  and $V$ is positive during inflation.
{\it We conclude that for flat \K\, potentials \rf{flat} all masses squared of the matter fields are positive and there are no tachyons.} Note, that the condition of flatness of the {\K} potential is sufficient for the absence of tachyons in the matter sector but not necessary. For models with a non-flat {\K} potential the stability of the matter sector is determined by equation \eqref{general}.

\subsection{Models with $g(\Phi)=0$}\label{gzero}
Another class of models that was studied in \cite{Kawasaki:2000yn,Kallosh:2010xz} has $g(\Phi)=0$. In this case the off-diagonal entries in the mass matrix in eq. \eqref{matrix} are
\be
c_1 A^g_{ij} + c_2 (A^g_{ij})'+ \bar f A^f_{ij} =(K^{\Phi\bar \Phi} \partial_\Phi k\, (\overline{g'+g \partial_\Phi k}) -\bar g) A^g_{ij} +  K^{\Phi\bar \Phi}\, (\overline{g'+g \partial_\Phi k}) (A^g_{ij})'+ \bar f A^f_{ij} = \bar f A^f_{ij}\,.
\ee
In these models we necessarily have $f\neq 0$ in order to have a non-trivial inflationary sector. If we choose however $A^f_{ij}=0$, i.e. there is no coupling of $S$ to the matter fields $U^i$ at the quadratic level in the superpotential in \rf{coupling}, then the off-diagonal entries in the mass matrix all vanish ($c=0$ in \eqref{matrix2}). The masses squared in the matter sector are then given by
\be
 \mu_{i}^{2}= V +\e^k \lambda_i^2 > 0\,.
\ee
So we see that this is another class of models without any tachyons in the matter sector during and after inflation.

\subsection{`Sgoldstino-less' models  with $c_2 (A^g_{ij})'+ \bar f A^f_{ij} \propto  A_{ij}^g$}\label{DZ}

Next we consider a particular case for which $c_2 (A^g_{ij})'+ \bar f A^f_{ij} \propto  A_{ij}^g$. Such a class of models was proposed by  in \cite{Dall'Agata:2014oka}, where they take the superpotential to be $W(\Phi, S, U^i) = f^{\rm DZ}(\Phi, U^i) (1+\sqrt 3 S)$. In this case we have
\be
f(\Phi) = \sqrt{3} g(\Phi)\,, \qquad W_{mat}^f= \sqrt{3} W^g_{mat}\,, \qquad \text{and} \qquad  f^{\rm DZ}(\Phi, U^i) = g(\Phi) + W^g_{mat}(\Phi,U^i)\,.
\ee
This implies
\be
\bar f A^f_{ij} = 3 \bar g A^g_{ij}\,.
\ee
Furthermore the authors imposed a $\mathbb{Z}_2$ symmetry of the scalar potential that amounts in our notation to $(A^g_{ij})'=0$. Lastly, the {\K} potential in the inflationary sector is taken to be canonical $k(\Phi,\bar \Phi) = -\frac12 (\Phi -\bar \Phi)^2$, so it satisfies the {\K} flatness condition in eq. \eqref{flat}. This leads to the following off-diagonal entries in the mass matrix
\be
c_1 A^g_{ij} + c_2 (A^g_{ij})'+ \bar f A^f_{ij} = (K^{\Phi\bar \Phi} \partial_\Phi k (\overline{g'+g \partial_\Phi k}) -\bar g) A^g_{ij} +  3 \bar g A^g_{ij} = 2 \bar g A^g_{ij}\,.
\ee
Therefore the eigenvalues in this case are given by (see eq. \eqref{general})
\be
 \mu_{i}^{2}= V + \e^{k} |g|^2 +\e^k \lambda_i \lp \lambda_i \pm 2|g|\rp = V + \e^{k} \lp \lambda_i \pm |g|\rp^2>0\,.
\ee
Thus we reproduced in our language the conclusion of \cite{Dall'Agata:2014oka} that these models have no tachyons in the matter sector. 

To obtain these results, we assumed that the field $S$ is either nilpotent, or very strongly stabilized  at $S = 0$, which can be achieved by adding higher order terms such as $(S\bar S)^{2}$ to the \K\ potential. Indeed, in the limit of strong stabilization, one can reproduce the results obtained in the theory of nilpotent fields \cite{Kallosh:2015pho}. 

One way to produce the  stabilizing terms in the \K\ potential is to  generate them by quantum effects.  Implementation of this method in  \cite{Dudas:2016eej} in application to the models of \cite{Dall'Agata:2014oka} required deviation from the rules formulated in \cite{Dall'Agata:2014oka}. Consequently, it was difficult to strongly stabilize the field $S$ and avoid tachyons in the models proposed in \cite{Dudas:2016eej}. However, it is not necessary to generate higher order terms in the \K\, potential by quantum effects. These terms are related to curvature invariants of the \K\ geometry, which can be large already at the tree level.

\section{Models with orthogonal nilpotent multiplets}\label{orthogonal}

Inflationary models based on an orthogonal nilpotent multiplet like the ones in \cite{Ferrara:2015tyn,Carrasco:2015iij,Dall'Agata:2015lek} provide another interesting class of models that can easily be coupled to matter. In these models the terms in the scalar potential that are linear or quadratic in $D_\Phi W = \partial_\Phi W +W \partial_\Phi K$ are absent  in the inflaton potential.  For $\Phi = \bar\Phi$, $S = 0$, $U^i=0$ the potential is 
\be
V = f^{2}(\Phi) - 3g^{2}(\Phi) \ ,
\ee 
and $K$ vanishes along the inflationary trajectory.
 In our language, for $U^i=0$ this amounts to setting terms proportional to $D_\Phi W = g'+g \partial_\Phi k$ equal to zero when we work with the general case in Sec. \ref{sec:general}. This does not change the diagonal entries in the mass matrix in eq. \eqref{matrix} but simplifies the off-diagonal entries as follows
\be
-\bar g  A^g_{ij} + c_2 (A^g_{ij})'+ \bar f A^f_{ij} = -\bar g A^g_{ij} + \bar f A^f_{ij} \,,
\ee
since
\be
c_1=K^{\Phi\bar \Phi} \partial_\Phi k \, (\overline{g'+g \partial_\Phi k}) -\bar g= -\bar g\,,
\ee
and 
\be
c_2= K^{\Phi\bar \Phi}\, (\overline{g'+g \partial_\Phi k})=0\,.
\ee
This means, for example, that a direct cubic coupling between matter and inflaton, $W_m\sim \Phi U^i U^j$, does not lead to instabilities  in these models. 
However, couplings to $S$, such as $W_m=  S A_{ij}^fU^i U^j$, may lead to instabilities.

If we now consider models where the matter fields $U^i$ do not couple to $S$ at the quadratic level, then $A^f_{ij}=0$. The mass matrix eigenvalues are then given by the same formula as in subsection \ref{Kflat}
\be
\mu_i^2= V + |g|^2 + \lambda_i \lp \lambda_i \pm | g|\rp= V + {3\over 4} |g|^2 + \Big |  \lambda_i \pm {1\over 2} | g|\Big |^2  > 0\,.
\ee
In this case there are also no tachyons in the matter sector during and after inflation. 
Note that this result is valid in this class of models even if $A^g_{ij}$ does depend on $\Phi$, i.e. even if $(A^g_{ij})' \not = 0$. This may simplify the construction of phenomenologically acceptable models of this type in cases where there is a direct interaction of matter fields with the inflaton in the superpotential. 

Another interesting feature of the models with orthogonal nilpotent multiplets is that the inflatino $\chi^\phi$ is proportional to the fermion of the $\BS$
multiplet $\chi^s$. Therefore in unitary gauge, where $\chi^s=0$ at the minimum of the potential, the true goldstino
\be
v= e^{K\over 2} \Big ( \chi^s D_S W  + \chi^\phi D_\Phi W \Big) 
\ee
vanishes:
\be
\chi^\phi \sim \chi^s \qquad \Rightarrow \qquad v|_{\chi^s=0}=0 \ .
\ee
The massive gravitino is decoupled from other spin 1/2 fermions and the analysis of the reheating is relatively simple. In models where the inflaton is not orthogonal to $\BS$, i.e. where  $\BS(\BPhi- \BPhibar)\neq 0$ one has to deal either with the mixing of the gravitino with the inflatino, $\gamma^\mu \psi_\mu \chi^\phi D_\Phi W$, in the $\chi^s=0$ gauge,
or in a gauge $v=0$ with complicated non-linear inflatino couplings. 

Thus, the models with orthogonal multiplets have simplified the fermion sector of the inflationary models. However, now when matter was added, the goldstino has to be supplemented by additional terms
\be
v= e^{K\over 2} \Big ( \chi^s D_S W  + \chi^\phi D_\Phi W  + \chi^{u^i} D_{U^i} W\Big) \,.
\ee
In principle, terms like that could destroy the simplicity of the fermion sector at the minimum since the gravitino is mixed with matter fermions $\chi^{u^i}$. Fortunately, our matter couplings are such that 
\be
D_{U^i} W= \partial _{U^i} W + K_{U^i} W\Big |_{U^i=0}=0\,.
\ee
Therefore the decoupling of the massive gravitino from all the spin 1/2 fermions of the theory is preserved at the minimum of the potential in models with orthogonal nilpotent multiplets in the unitary gauge $\chi^s=0$.

\section{Examples without tachyons}\label{examples}

\subsection{A simple model with a single matter multiplet}

The example below is one of the first and simplest models where one can confirm stability of the inflationary trajectory and absence  of tachyons when coupling the inflationary sector to matter. We considered  nilpotent and orthogonal superfields, $\BS^2=\BS (\BPhi- \BPhibar)=0$,  which further simplifies the study. The model is
\ba\label{KdiskNewDiskM}
K&=& -{3\over 2} \alpha \log \left[{(1- \Phi\bar\Phi)^2\over (1-\Phi^2) (1-\bar \Phi^2)}  \right]+S\bar S +  U\bar U \,,\cr\cr
W &=&  g(\Phi) + S f(\Phi) +  {m\over 2} U^2 \,,
\ea
where we can take $m>0$ by absorbing its phase in $U$. The mass eigenvalues of the complex matter field $U$ are then
\be\label{mi}
 \mu^{2} =V+ |g|^2 \pm  |g|m + m^{2}=   V+{3g^{2}\over 4} + \left(m \pm {|g|\over 2} \right)^{2} \geq 3\left(H^2+{g^{2}\over 4}\right),
\ee
confirming the general case in eqs. \rf{masses}, \rf{Eigenvalues*} with $\lambda_1= m$. The last term in this equation takes into account that $V \approx 3H^{2}$ during inflation.
 
One can also check that, in agreement with our general discussion, the mass eigenvalues remain positive if  $g(\Phi) = const$ and one adds a term proportional to $\Phi U^2$ to the superpotential.

Whereas matter fields are already stable in this model, it is interesting  to check what will happen, if one adds a higher order term $-\zeta S\bar S U\bar U$  to the \K\ potential. The result is that it adds $\zeta |f(\Phi)|^2$ to the mass eigenvalues (\ref{mi}), which provides additional stabilization of the fields for $\zeta > 0$, as explained for the general case in Sec. \ref{sec:curv}.

\subsection{A model with two matter multiplets}

Our next example has generic functions $f(\Phi)$ and $g(\Phi)$ in the superpotential \rf{eq:inf} and the hyperbolic geometry inflaton independent \K\, potential in \rf{KdiskNewDisk}.  We couple this inflationary model with  matter fields $U$ and $Y$, which we introduce according to the explanation above 
\ba\label{KdiskNewDiskM2}
K &=&  -{3\over 2} \alpha \log \left[{(1- \Phi\bar\Phi)^2\over (1-\Phi^2) (1-\bar \Phi^2)}  \right]+S\bar S + U\bar U +Y\bar Y\,,\cr\cr
W &=&  g(\Phi) + S f(\Phi) + {M\over 2} U^2 + {m\over 2} UY\,,
\ea
where we can take $m,M>0$ without loss of generality. Here the inflaton superfield $\Phi$ is a disk variable, $\Phi\bar \Phi <1$, $S$ is a nilpotent superfield, both belong to the inflaton sector of the theory. The inflaton is $\Phi+\bar \Phi$ in these models where $\Phi-\bar \Phi=0$ during inflation.

The two matter multiplets, $U$ and $Y$ have canonical \K\, potentials and quadratic dependence in $W$. In this relatively simple case, we can compute the mass eigenvalues for the matter fields $U$ and $Y$ either directly, or using the general formula \rf{Eigenvalues}. These two computations agree,  the relevant values of $\lambda_1$ and $\lambda_2$ being $\sqrt {m^2+ M^2}\pm M$. As before, all 4 eigenvalues of the mass matrix are greater than $V \approx 3H^{2}$ during inflation:
\be
 \mu_{i}^{2} =V+ {3|g|^2\over 4} +\left(\sqrt{M^{2}+m^{2}} \pm M \pm {|g|\over 2}\right)^{2},
\ee
where $i = 1,2,3,4$ correspond to  four different combinations of the signs $\pm$ in this equation.

\vskip 5pt

\section{ Conclusions }

In this paper we considered some of the most popular  models of chaotic inflation in supergravity with \K\ potential with a flat direction, including general models with a canonical \K\ potential \cite{Kawasaki:2000yn,Kallosh:2010xz}, as well as the advanced version of $\alpha$-attractor models  \cite{Carrasco:2015uma,Carrasco:2015pla,Carrasco:2015rva}, based on the hyperbolic geometry of the moduli space \cite{Kallosh:2015zsa}. Flatness of the \K\ potential in these models helps to ensure flatness of the inflaton potential. In this paper we have shown that this class of models has an additional advantage: One can easily add matter fields to these models without destabilizing the inflationary trajectory and even without affecting the inflationary evolution at all. This means, in particular, that under certain conditions outlined in this paper, one can add matter fields without creating tachyonic instabilities or forcing many fields to evolve simultaneously: Inflation may remain driven by a single field even if many interacting matter  fields are present.

The simplest version of the models protected from tachyons has the following features: The matter has no direct coupling to the inflationary sector in the \K\, potential and  in the superpotential.\footnote{Note that the condition that  matter has no direct coupling to the inflationary sector in the \K\, potential and  in the superpotential is often used in supergravity cosmology, since under this condition the decay rate of the inflaton field to matter during reheating is strongly suppressed. This leads to a smaller value of the reheating temperature, which simplifies the solution of the cosmological gravitino problem \cite{Ellis:1982yb}.} 
 In terms of the superpotential in eq. \rf{coupling} it means that $\partial_\Phi A^g_{ij}(\Phi) =0$ and $A^f_{ij}(\Phi)=0$.
The matter part of the superpotential has to start with terms quadratic in matter fields or higher, as  in eq. \rf{quadratic}. Finally the models have to have a flat \K\, potential, strictly independent of the inflaton direction, as  in eq. \rf{flat}. When all three conditions are satisfied, one can guarantee that the matter sector will not destabilize an underlying successful model of inflation, which also means that there are no tachyons. The mass eigenvalues for these models  during inflation are derived in eq. \rf{Eigenvalues}, which can also be given in the form
\be
\mu_i^2 = V + {3\over 4} |m_{3/2}(\Phi)|^2 + \Big |  \lambda_i \pm {1\over 2} | m_{3/2}(\Phi)|\Big |^2  > 0\,,
\label{Eigenvalues1}\ee
where $V\approx 3H^2$ during inflation, $m_{3/2}(\Phi) \equiv g(\Phi)$ is the inflaton-dependent mass of the gravitino, and the $\lambda_i$ are related  to  the $\Phi$-independent coupling $A_{ij}$ in eq. \rf{coupling}.  Since all mass eigenvalues are greater than $3H^2$, all matter fields quickly reach their minima at $U^i=0$ during inflation.

We have also explained in sec. \ref{curvature} why the moduli space curvature presents a universal mechanism for moduli stabilization. Such a curvature creates a wall in the potential for the moduli which we would like to restrict to vanishing values. In principle,  the curvature can make these moduli fields stable and heavy.  This mechanism can be used in addition to other stabilization tools which we studied in this paper.

We stress that our conditions in models with independent $f(\Phi)$ and $g(\Phi)$ in the inflationary superpotential are sufficient for a successful coupling of the inflationary sector to matter, but they may not be necessary. For example, in models with flat \K\, potential the decoupling requirement between the inflationary sector and matter in the superpotential may be relaxed without affecting our main result for the positivity of masses: we still need that in eq. \rf{quadratic} $\partial_S  A_{ij} =0$ which means $A_{ij}^f=0$ in eq. \rf{coupling}. Also we need that either $\partial_\Phi  A_{ij} =0$ or $g'(\Phi)=0$.
But cubic terms with $  B_{ijk} $ in eq. \rf{quadratic} may depend on the inflationary sector fields, and the same for higher terms in the $U^i$. Stabilization in some special models, for example, the ones with $g(\Phi)=0$, was presented in sec. \ref{gzero}.
Some other cases that were studied in \cite{Dall'Agata:2014oka} and for which $  f(\Phi)= \sqrt 3 g(\Phi)$ where also argued to lead to a successful coupling to matter, under different conditions. Here we explained these results as a particular case of our general argument with details given in sec. \ref{DZ}. The simplified features of matter moduli stabilization in models with orthogonal nilpotent multiplets \cite{Ferrara:2015tyn,Carrasco:2015iij,Dall'Agata:2015lek}  are given in sec. \ref{orthogonal}. Various successful examples of stabilization are given in sec. \ref{examples}. Examples of  unsuccessful models, violating the conditions of our theorems, can be found in appendices A.1 and A.2.

To develop these results towards a more realistic model, like combining the MSSM or NMSSM models with an inflationary model, we will note here, as also noticed in \cite{Dall'Agata:2014oka} in the context of their models, that we can now introduce, for example, the squark and slepton multiplets with vanishing vev's.  Note that tachyonic instability in the models violating some of our conditions may play a constructive role,  creating  non-vanishing vev's, see appendix \ref{higgs}. Further investigation  of such models will appear in a separate publication \cite{neutrino}.

\

We are grateful to J. J. Carrasco, S. Ferrara, D. Roest and J. Thaler for  enlightening  discussions and collaboration on related projects.   The work of RK  and AL is supported by the SITP, and by the NSF Grant PHY-1316699.  The work of AL is also supported by the Templeton foundation grant `Inflation, the Multiverse, and Holography'. The work of TW is supported by COST MP1210. TW thanks the Department of Physics of Stanford University for the hospitality during a visit in which this work was initiated.

\appendix

\section{Examples with tachyons}\label{app:examples}

\subsection{Tachyons in models with $S$-coupled matter}

Let us start out by looking at an example in which we satisfy all the conditions except the decoupling of the matter sector from $S$  in the superpotential. In particular, we take the {\K} and superpotential
\ba
K &=&  -{3\over 2}   \alpha \log \left[{(1- \Phi\bar\Phi)^2\over (1-\Phi^2) (1-\overline \Phi^2)}  \right] +S\bar S + U \bar U\,,\cr\cr
W&=& S f(\Phi) + g(\Phi) +m S U^2\,.
\ea
The term in \rf{Scoupling} is present and in this case we do not expect a protection from tachyons.
 It is straightforward to calculate the masses squared for the field $U$ during inflation and we find
\be\label{eq:masses}
\mu^2 = V \pm 2 |m f| + |g|^2\,.
\ee
We see that the positivity of the masses squared is not guaranteed and in particular at the end of inflation we have $V \approx 0$ and $ |g|^2 = m_{3/2}^2$, the gravitino mass squared. Thus we see from eq. \eqref{eq:masses} that in this model it is impossible to make both masses for $U$ larger than the gravitino mass  in the state with $U = 0$.

\subsection{Tachyons in models with a non-flat \K\, potential}

Consider the early versions of $\alpha$-attractor models \cite{Kallosh:2013hoa} which in half-flat variables can be presented as follows
\ba \label{halfplane}
&&K= -{3}   \alpha \log \left[\Phi+ \bar\Phi  \right] +S\bar S\,,  \cr
\cr
&&W= \Phi^{\frac{3\alpha}{2}} \Big( S f(\Phi) + g(\Phi)\Big)\,.
\ea
Here $K_\Phi \neq 0$ at $\Phi=\bar \Phi$, i. e. the \K\, potential is not flat during inflation.
 
Now we add  the interaction with matter according to our rules in \rf{Wdecoupled} with decoupled matter in $W$
 \ba
&&K= -{3}   \alpha \log \left[ \Phi+ \bar\Phi  \right] +S\bar S +  U\bar U \, , \cr
\cr
&&W= \Phi^{\frac{3\alpha}{2}} \Big( S f(\Phi) + g(\Phi)\Big) + {M\over 2} U^2 \ .
\ea
We can compute the masses of the complex field $U$ and, taking $M>0$, we find the eigenvalues of the mass matrix during inflation to be
\be
\mu^2 =  V + \e^{k} |\tilde{g}|^2 +\e^k  M \lp  M \pm |(\Phi+\bar \Phi) \tilde{g}'+(1-3\alpha) \tilde{g}|\rp\,,
\ee
with $\tilde{g}(\Phi) = \Phi^{\frac{3\alpha}{2}} g(\Phi)$. One can check that the eigenvalue with the minus sign is not positive definite so the general formula \rf{general} allows for tachyons in the matter sector, i.e. it cannot be excluded in general that the inflaton sector destabilizes the matter fields.

Now reconsider the same model by performing a \K\, transformation of the kind made in \cite{Carrasco:2015uma}. This leads to 
 \ba
&&K= -{3\over 2 }   \alpha \log \left[{( \Phi+ \bar\Phi)^2\over 4 \Phi \bar \Phi}  \right] +S\bar S +  U\bar U \, , \cr
\cr
&&W=  \Big( S f(\Phi) + g(\Phi)\Big) +   {M\over 2} \Phi^{-\frac{3\alpha}{2}} U^2 \ .
\ea
This model has a flat \K\, potential but it also has a direct coupling between the inflaton and the matter in the superpotential. In such case, the flatness of the \K\, potential is not sufficient to protect the matter field $U$ against becoming tachyonic due to the presence of the inflationary sector (cf. eqs. \eqref{eq:1}, \eqref{eq:2}). If, on the other hand, one  makes the simplest choice
\ba
&&K= -{3\over 2 }   \alpha \log \left[{( \Phi+ \bar\Phi)^2\over4  \Phi \bar \Phi}  \right] +S\bar S +  U\bar U \,, \cr
\cr
&&W= S f(\Phi) + g(\Phi) +  {M\over 2} U^2\,,
\ea
then one finds that the masses are positive and given by eq. \eqref{Eigenvalues} with $\lambda=M$.

\subsection{Models with tachyons and stable Minkowski vacuum with symmetry breaking}\label{higgs}

In this paper we concentrated on models where one can avoid tachyons during inflation. But tachyons are not always bad, as they are often responsible for spontaneous symmetry breaking. Also, hybrid inflation is an example of inflationary models where a tachyonic instability can be used constructively \cite{Linde:1991km}. 

As a simplest example, consider the  model
\ba \label{mink}
K &=&  -{3\over 2} \alpha \log \left[{(1- \Phi\bar\Phi)^2\over (1-\Phi^2) (1-\bar \Phi^2)}  \right]+S\bar S + U\bar U \ ,  \cr \cr
W &=&  S \left(\Phi^{2}+ \lambda\big(U^{2}- {c^{2}} \big)\right) \, .
\ea
Writing $\Phi = \phi = \tanh{\varphi\over\sqrt{6\alpha}}$, $U = u+ i \, v$, one finds the potential
\be
V =e^{u^{2}+v^{2}}\left( \left(\phi^{2}+\lambda (u^{2}-v^{2}-c^{2})\right)^{2}  + 4 \lambda^{2} u^{2} v^{2}\right) \ ,
\ee
where one can check that $\text{Im}(\Phi)=0$ is a minimum for  $u, v =0$. Note that this expression is always positive or zero. We will assume that $c< 1$, $\lambda < 1/3$. Then during inflation at $\phi \approx 1$ the potential has a stable dS minimum for the matter fields at $u = v = 0$, but also two deep separate Minkowski minima at large $v$ and $u = 0$. With a decrease of $\phi$ from $1$ to $\sqrt\lambda\sqrt{2+c^{2}}$, the minimum at $u = v = 0$ becomes unstable, and the system may (if it has enough time) fall to one of the two minima at $u = 0$, $v \not = 0$. However, during the subsequent decrease of the field $\phi$ below $\sqrt\lambda\, c$, these two minima merge, and two other minima appear, at $v = 0$, $u \not = 0$. In the end of the process the fields fall into one of the two supersymmetric Minkowski minima $\phi = v = 0$, $u = \pm c$.

Thus we see that the post-inflationary cosmological evolution can be very complicated, with two different stages of tachyonic instability, but in the end the universe falls into one of the two Minkowski minima.

Of course, this is just a toy model, and several things should be done to make it realistic. First of all, in the process of spontaneous symmetry breaking, the universe becomes divided into domains with $u = \pm c$ separated by domain walls, which is a cosmological disaster unless $c$ is extremely small. This can be avoided in more complicated models; it is known for example that domain walls do not appear after symmetry breaking in the standard model. Yet another issue is that SUSY is unbroken in the minima $\phi = v = 0$, $u = \pm c$, and it is not obvious whether one can break it without making these minima AdS, which is undesirable.  Therefore this toy model should be modified to address these issues.  However, this example indicates that tachyonic instabilities during or after inflation could be useful for describing spontaneous symmetry breaking. We hope to return to a discussion of this possibility in a separate publication \cite{neutrino}.

\end{document}